
\def\Z_2{{Z \!\!\! Z}_2}
\def\[{[\![}
\def\]{]\!]}

\baselineskip=18pt
\leftskip 36pt

\hfill INRNE-TH-94/3
\vskip 18pt
\noindent
{\bf Wigner quantum oscillators. Osp(3/2) oscillators}

\vskip 32pt
\noindent
T. D. Palev\footnote*{Permanent address: Institute for Nuclear Research
and Nuclear Energy, Boul. Tsarigradsko Chausse 72,
1784 Sofia, Bulgaria; E-mail
palev@bgearn.bitnet}

\noindent
Arnold Sommerfeld Institute for Mathematical Physics,
Technical University of Clausthal, 38678
Clausthal-Zellerfeld, Germany

\vskip 12pt
\noindent
N. I. Stoilova

\noindent
Institute for Nuclear Research and Nuclear Energy, 1784
Sofia, Bulgaria

\vskip 32pt
{\bf Abstract.}
The properties of the three-dimensional noncanonical
$osp(3/2)$ oscillators, introduced in [J. Phys.  A: Math.
Gen.  {\bf 27} (1994) 977], are further studied.  The
angular momentum $M$ of the oscillators can take at most
three values $M=p-1,p,p+1$, which are either all integers
or all half-integers. The coordinates anticommute with each
other.  Depending on the state space the energy spectrum
can coincide or can be essentially different from those of
the canonical oscillator.  The ground state is in general
degenerated.

\vskip 48pt

\noindent
{\bf 1. Introduction }

\noindent
In the present paper we continue the study  of
the three-dimensional nonrelativistic quantum $osp(3/2)$
oscillators, introduced in [1]. The main algebraic feature
of each such oscillator is that its position and momentum
operators generate a representation of the orthosymplectic
Lie superalgebra $osp(3/2)$.  The state space of each
oscillator is an infinite-dimensional irreducible module of
the Lie superalgebra (LS) $osp(3/2)$. This result was only
announced in [1]. Here we prove it.  Moreover in [1] we
have considered  one particular oscillator, namely the one
with an angular momentum $1/2$, stating  only the energy
and the angular momentum spectrum of the other possible
oscillators. Here we  study the physical properties of all
$osp(3/2)$ oscillators in detail, introducing an
orthonormed basis, consisting of common eigenvectors of
the Hamiltonian $H$,  the square of the angular momentum
${\bf M}^2$ and its third projection $M_3$.
Within each state space we compute  the matrix elements of
essentially all physical observables.

The motivation to introduce and study such more general,
noncanonical oscillators was outlined in [1]. We recall the
main points.  The idea belongs to Wigner [2], who observed
that the Hamiltonian equations can be identical with the
Heisenberg equations for position and momentum operators,
which do not necessarily satisfy the canonical commutation
relations (CCRs). Wigner has demonstrated this on an
example of an one-dimensional harmonic oscillator, studied
subsequently by several authors [3].

The question about the compatibility of the Hamiltonian equations

$$\dot{\bf p}=-m \omega^2 {\bf r}, \quad \dot{\bf r}={{\bf
p}\over m} \eqno(1) $$

\noindent
with the Heisenberg equations (here and throughout
$[x,y]=xy-yx$, $\{x,y\}=xy+yx)$

$$\dot {\bf p}=-{i\over \hbar}[{\bf p},H], \quad
\dot {\bf r}=-{i\over \hbar}[{\bf r},H], \eqno(2)  $$

\noindent
of a three-dimensional harmonic oscillator, namely of a
system with a Hamiltonian

$$H={{\bf p}^2\over 2m}+
    {m \omega^2 \over 2}{\bf r}^2, \eqno(3)  $$

\noindent
was investigated in [4] and [1]. The present paper is also
in the same frame. The problem is to determine and study at
least some noncanonical Wigner quantum oscillators. To be
more precise with the terminology we give the following
definition.

{\it Definition 1.} A triple ({\bf r, p}, $W$) is said to
be a Wigner quantum oscillator, if it fulfills the
following three conditions (we refer to them as to quantum
conditions):

\smallskip

(i) The state space of the oscillator $W$ is a Hilbert
    space. The physical observables are Hermitian (linear)
    operators in $W$.

\smallskip
(ii) The Hamiltonian equations (1) and the Heisenberg
     equations (2) are identical (as operator equations) in
     $W$.

\smallskip
(iii) The projections of the angular momentum of the oscillator
      {\bf M}=$(M_1, M_2, M_3)$ are in the
      enveloping algebra of the position operators
      {\bf r}=$(r_1, r_2, r_3)$ and the momentum operators
      {\bf p}=$(p_1, p_2, p_3)$. Each $M_k$ is linear
      in $(r_1, r_2, r_3)$ and linear in $(p_1, p_2, p_3)$,
      so that {\bf M}, {\bf r} and {\bf p} transform as
      vectors:

      $$[M_j,c_k]=i \sum_{l=1}^3 \varepsilon_{jkl}c_l,
            \quad c_k=M_k,r_k,p_k,\; j,k=1,2,3 \eqno(4) $$

We underline that in our approach the operators
{\bf r}=$(r_1, r_2, r_3)$ and {\bf p}=$(p_1, p_2, p_3)$ are
postulated to be the position and the momentum operators of
the oscillator, independently of the fact that they do not
satisfy the CCRs.

The mathematical problem that arises is to find the unknown
operators {\bf r}=$(r_1, r_2, r_3)$ and
{\bf p}=$(p_1, p_2, p_3)$ so that the quantum conditions
(i)-(iii) hold. To this end it is convenient to pass to new
unknown operators

$$a_k^\pm=\sqrt{m \omega \over 2 \hbar} r_k \mp
{i \over \sqrt {2m \omega \hbar}}p_k, \quad k=1,2,3. \eqno(5)  $$

\smallskip
\noindent
For the sake of convenience we refer to the operators
$a_1^\pm$, $a_2^\pm$, $a_3^\pm$ as to creation and
annihilation operators (CAOs). These operators should not
be confused with the Bose operators.  They  are unknown
operators we are searching for. Only in one particular
representation, corresponding to the canonical case, (5)
are Bose CAOs.  In terms of $a_k^\pm$ the Hamiltonian (3)
reads:

$$H={1 \over 2}\omega \hbar \sum_{k=1}^3 \{a_k^+, a_k^- \}.\eqno(6)  $$

\noindent
The condition (ii) yields ($k=1,2,3$):

$$\sum_{i=1}^3 [ \{a_i^+,a_i^- \},a_k^\pm]=\pm 2a_k^\pm
                                            \eqno(7)  $$

\noindent
and it is equivalent to the requirement that the
Hamiltonian (in units $\omega \hbar)$, namely $N=(\omega
\hbar)^{-1}H$ is a number operator,

$$[N,a_k^\pm]=\pm a_k^\pm. $$

The equations (7) are a unique consequence from the
Hamiltonian equations (1) and the Heisenberg equations (2)
independently of the properties of the unknown CAOs
$a_k^\pm$. They are equal time relations, the time
dependence being

$$a_k^\pm(t)=e^{\pm i
\omega t}a_k^\pm,
\quad a_k^\pm (0) \equiv a_k^\pm ,
\quad k=1,2,3 . \eqno(8) $$

\noindent
Hence eqs. (7) hold, if they are fulfilled at, say, $t=0$.

Let $F$ be the free unital (=with unity) associative
algebra with generators $a_1^\pm$, $a_2^\pm$, $a_3^\pm$ and
relations (7).  Any representation of $F$ is a candidate
for a Wigner quantum oscillator.  Out of all such
representations one has to select those for which
also  conditions (i)-(iii) hold.  The set of the solutions
is not empty, it contains at least the canonical oscillator
solution.  The general solution of the problem is however
unknown to us. In [1] we have listed three classes of
solutions of CAOs fulfilling the compatibility eqs. (7). The
third class of operators $a_1^\pm,a_2^\pm,a_3^\pm$ are the
creation and annihilation operators of the $osp(3/2)$
oscillator. They are defined with the following relations
($\varepsilon , \xi=\pm \;{\rm or}
\pm 1, \; i,j,k=1,2,3$):

$$\vcenter{\openup3\jot \halign{$#$ \hfil  \cr
[\{a_i^+,a_j^-\},a_k^\varepsilon]={2 \over 3}\delta_{ik}a_j^\varepsilon
-{2 \over 3}\delta_{jk}a_i^\varepsilon +
{2 \over 3}\delta_{ij}\varepsilon a_k^\varepsilon, \cr
[\{a_i^\varepsilon,a_i^\varepsilon \},a_k^{\varepsilon}]=0, \cr
\{a_i^\varepsilon,a_j^\varepsilon \}=0,\; i \not= j, \cr
\{a_i^+,a_j^-\}=-\{a_j^+,a_i^-\},\; i \not= j, \cr
\{a_1^\varepsilon,a_1^\xi\}=\{a_2^\varepsilon,a_2^\xi\}
=\{a_3^\varepsilon,a_3^\xi\},\cr
}} \eqno(9)$$

In the next section we investigate the algebraic structure of
the operators (9) and establish their relation to the
$osp(3/2)$ algebra.  To this end we first recall the
definition of the Lie superalgebra $osp(3/2)$.

\vskip 8mm
\noindent
{\bf 2. Lie superalgebraic properties of the creation and the
annihilation operators (9)}

\bigskip
\noindent
In a matrix form the orthosymplectic LS $osp(3/2)$  can be
defined as the set of all 5-by-5 matrices of the form [5]

$$\pmatrix{a & 0  &  b & \vert &  x &  u \cr
           0 & -a &  c & \vert &  y &  v \cr
	  -c & -b &  0 & \vert &  z &  w \cr
	  -- & -- & -- &       & -- & -- \cr
	   v &  u &  w & \vert &  d &  e \cr
	  -y & -x & -z & \vert &  f & -d \cr}, \eqno(10)  $$

\noindent
where the nonzero entries are arbitrary complex numbers.
The even subalgebra $so(3) \oplus sp(2)$ consists of all
matrices (10), for which $x=y=z=u=v=w=0$, whereas the odd
subspace is obtained taking $a=b=c=d=e=f=0$. The product (=
the supercommutator) $\[\;,\;\]$ of any two homogeneous
elements is (a) a matrix anticommutator between odd
matrices and (b) a matrix commutator in all other cases.

The algebra $osp(3/2)$ is generated from its subspace $G$,
consisting of all matrices (10), for which
$a=d=e=f=x=y=u=v=0$ [6]. Let $e_{ij}$ be a 5-by-5 matrix
with 1 on the cross of the $i^{th}$ row and the $j^{th}$
column and zero elsewhere. The matrices

$$c_0^-=\sqrt{2}(e_{23}-e_{31}), \quad
  c_0^+=\sqrt{2}(e_{32}-e_{13}), \eqno(11)$$

\noindent

$$c_1^-=\sqrt{2}(e_{34}-e_{53}), \quad
  c_1^+=\sqrt{2}(e_{35}+e_{43}), \eqno(12)$$

\noindent
constitute a basis in $G$ with even generators (11) and
odd generators (12). The other 8 $osp(3/2)$ generators are
the supercommutators of (11)-(12),

$$ \[c_p^\xi,c_q^\eta \]=c_p^\xi c_q^\eta
-(-1)^{pq}c_q^\eta c_p^\xi,
\quad \xi, \eta =\pm, \quad p,q
\in \Z_2 \equiv (0,1). \eqno(13)  $$

The subspace $G$ is a Lie-super triple system in the
terminology of Okubo [7]. It is closed under double supercommutators,

$$\[\[x,y\],z\]=2<y \vert z>x-2(-1)^{deg(x)deg(y)}<x \vert
z>y \in G,\; \forall x,y,z \in G, \eqno(14) $$

\noindent
where the bilinear form  $<x \vert y>$ is defined as [7]

$$<c_p^\xi \vert c_q^\eta >=\eta^p \delta_{pq}\delta_{\xi,-\eta},
\quad \xi, \eta =\pm, \quad p,q
\in \Z_2 \equiv (0,1). \eqno(15)  $$

In terms of the basis (11)-(12) in $G$ eq. (14) reads:

$$\[\[c_p^\xi,c_q^\eta \],c_r^\varepsilon \]=
2\varepsilon^r \delta_{qr}\delta_{\varepsilon,-\eta}c_p^\xi-
2\varepsilon^r (-1)^{qr}\delta_{pr}\delta_{\varepsilon,-\xi}
c_q^\eta,
\quad \xi, \eta , \varepsilon =\pm, \quad p,q,r
\in \Z_2 . \eqno(16)  $$

\noindent
In another form the eq. (16) was derived in [6]. There it
was shown that $B^\pm =c_1^\pm$ are para-Bose operators [8],
whereas the operators $F^\pm =c_0^\pm$ are para-Fermi
operators [8]. Observe that the para-Fermi
operators appear as even (i.e. bosonic) variables, whereas the
parabosons are odd (i.e. fermionic) operators. Moreover
the parabosons do not commute with the parafermions. To
the same conclusion arrived recently also Okubo [7] and
Macfarlane [9].  It may be interesting to observe that the
eqs. (16) are satisfied with ordinary bosons and fermions,
provided that the bosons anticommute with the fermions [6].
In this way one obtains the simplest infinite-dimensional
representation of $osp(3/2)$.

The eqs. (16) were derived using the 5-dimensional
representation (10) of $osp(3/2)$. Since however during the
derivation we were using only supercommutation relations,
eqs. (16) hold within every representation. Therefore from
now on (without changing the notation) we consider $c_p^\xi$
($p=0,1$; $\xi=\pm$) as abstract, representation independent
generators. It is essential to point out that the  supercommutation
relations between all  generators $c_p^\xi, \;
\[c_p^\xi,c_q^\eta \]=c_p^\xi c_q^\eta -(-1)^{pq}c_q^\eta
c_p^\xi \; (\xi, \eta =\pm; p,q= 0,1)$  can be
computed using only eqs. (16).  Therefore from the very
definition of an universal enveloping algebra we draw our
first conclusion.

\noindent
\smallskip
{\it Proposition 1.} The free associative unital
algebra $F_c$ of the paraoperators $c_p^\xi$  ($p=0,1$;
$\xi=\pm$) and the relations (16) is the universal
enveloping algebra $U[osp(3/2]$ of the Lie superalgebra
$osp(3/2)$. The $\Z_2$ grading on $U[osp(3/2]$ is induced
from the requirement that $c_0^\pm$ are even generators,
whereas $c_1^\pm$ are odd generators.

Define the following 6 elements from $F_c$:

$$a_1^\varepsilon={1\over {2 \sqrt{3}}}[c_1^\varepsilon,c_0^- -c_0^+],
\quad a_2^\varepsilon={i\over {2 \sqrt{3}}}[c_1^\varepsilon,c_0^- +c_0^+],
\quad a_3^\varepsilon={1\over {\sqrt{3}}}c_1^\varepsilon, \quad
\varepsilon=\pm. \eqno(17) $$

\noindent
It is straightforward to check that the operators (17)
satisfy the relations (9). Let $F_3(3)$ be the associative
subalgebra of $F_c$, generated by the odd elements
$a_1^\pm$, $a_2^\pm$, $a_3^\pm$,  $F_3(3) \subset F_c=U[osp(3/2]$.
Using only the relations (9), one derives

$$c_0^\varepsilon={3\over 2}[\varepsilon \{a_1^-,a_3^+ \}+
                            i\{a_2^-,a_3^+ \}]=
               {3\over 2}[-\varepsilon \{a_1^+,a_3^- \}-
                            i\{a_2^+,a_3^- \}], \quad
c_1^\varepsilon = {\sqrt 3}a_3^\varepsilon , \quad
\varepsilon =\pm. \eqno(18) $$

\noindent
Hence the operators $a_1^\pm$, $a_2^\pm$, $a_3^\pm$
generate the algebra $F_c$,

$$F_3(3)=F_c=U[osp(3/2]. \eqno(19) $$

\noindent
Thus, we have proved the result, announced in [1], namely

\noindent
\smallskip
{\it Proposition 2.} The free associative unital algebra
$F_3(3)$ of the CAOs $a_1^\pm$, $a_2^\pm$, $a_3^\pm$ and the
relations (9) is the universal enveloping algebra
$U[osp(3/2]$ of the Lie superalgebra $osp(3/2)$.  The $\Z_2$
grading on $U[osp(3/2]$ is induced from the requirement
that the creation and the annihilation operators are odd
elements.

\smallskip
$U[osp(3/2]$ can certainly be viewed as a Lie superalgebra
with a supercommutator defined as in every associative
superalgebra, namely $\[x,y\]=xy-(-1)^{deg(x)deg(y)}yx$.
The linear envelope of $c_p^\xi,\;
\[c_p^\xi,c_q^\eta \], \;\; \xi, \eta =\pm, \; \; p,q=0,1 $
is then a Lie subalgebra of the LS $U[osp(3/2]$ isomorphic to
$osp(3/2)$. From (17) and (18) one concludes that
(in the category of the Lie superalgebras)
the CAOs $a_1^\pm$, $a_2^\pm$, $a_3^\pm$ generate the subalgebra
$osp(3/2)$ of the LS $U[osp(3/2]$. More precisely,

$$lin. env. \{ \{a_i^\xi,a_j^\eta \},a_k^\varepsilon \vert
\xi , \eta ,  \varepsilon =\pm, \;
i,j,k=1,2,3 \}=osp(3/2). \eqno(20)  $$

\vskip 8mm

\noindent
{\bf 3. Satisfying the quantum conditions }

\bigskip
\noindent

\bigskip
\noindent
In view of the results of  Sec. 2 we already know that any
state space $W$ of the CAOs (9) is an $osp(3/2)$ module
(=representation space of the LS $osp(3/2)$). The problem
is to select those modules, for which the quantum conditions
(i)-(iii) hold.

\vskip 8mm

\noindent
{\bf Condition (ii)}

\bigskip
\noindent
Let $W$ be any $osp(3/2)$) module, i.e., a representation
space where the $osp(3/2)$ creation and annihilation
operators (9) are defined as linear operators. From eq. (8)
one obtains:

$$r_k(t)={\sqrt{\hbar \over 2m\omega }}
[a_k^+ e^{i \omega t}+a_k^- e^{-i \omega t}],\quad
p_k(t)=i\sqrt{m \omega \hbar \over 2}
[a_k^+ e^{i \omega t}-a_k^- e^{-i \omega t}]. \eqno(21) $$

\noindent
It is straightforward to check that the Hamiltonian
equations (1) and the Heisenberg equations (2) hold and
are identical as operator equations. Hence condition (ii) puts
no restriction on the $osp(3/2)$ modules, it holds within
each such module.
Already now we can say, that if the
operators (21) fulfilling all quantum conditions
 exist, then they possess quite unusual
properties. In particular from (9) and (21) one derives
that the different coordinates (resp. the different
momenta) of the oscillator anticommute:

$$\{r_i,r_j\}=\{p_i,p_j\}=0 \quad \forall i\neq j=1,2,3.
\eqno(22)$$

\vskip 8mm

\noindent
{\bf Condition (iii)}

\bigskip
\noindent
Consider the operators (21), defined as linear operators in
an arbitrary $osp(3/2)$) representation space $W$. Then the
projections $M_1$, $M_2$ and $M_3$ of the angular momentum
can be defined to be:

$$M_i=-{3\over 4\hbar}\sum_{j,k=1}^3 \varepsilon_{ijk}
\{r_j,p_k\}  \quad i=1,2,3. \eqno(23)  $$

In order to check that eqs. (4) hold it is better to express
the angular momentum components in terms of the CAOs (9),
namely

$$M_j=-{3i\over 4}\sum_{k,l=1}^3 \varepsilon_{jkl}
\{a_k^-,a_l^+\} \quad j=1,2,3, \eqno(24)  $$

\noindent
or in terms of the Lie-super triple generators $c_p^\xi$
($p=0,1$; $\xi=\pm$)

$$M_1=-{1\over 2}(c_0^+ +c_0^-), \quad
M_2={i\over 2}(c_0^+ -c_0^-), \quad
M_3={1\over 2}[c_0^+,c_0^-]. \eqno(25) $$

The angular momentum projections $M_1$, $M_2$, $M_3$ are the
generators of the $so(3)$ part of the even subalgebra of
$osp(3/2)$, sitting in the left upper corner of the matrix
realization (10).  Eqs. (25) give the usual realization of
$so(3)=sl(2)$ in terms of para-Fermi (and hence also in
terms of Fermi) operators.

\vskip 8mm

\noindent
{\bf Condition (i)}

\bigskip
\noindent
So far we have satisfied the quantum conditions (ii) and
(iii). These conditions put no restriction on the
representation space, they hold within each $osp(3/2)$
module. Passing to the condition (i), we have the first
restriction.

\smallskip
\noindent
{\it Proposition 3.} If the $osp(3/2)$ module W, satisfying
(i), exists then it is completely reducible.

\smallskip
The proof is standart. Indeed, assume that $E\subset W $ is
a subspace of the Hilbert space $W$, which is invariant
with respect to the Hermitian operators $a \in
(r_1, r_2, r_3, p_1, p_2, p_3)$. Denote by $F$  its
orthogonal compliment, $W=E \oplus F$. Let $ e_1, \ldots,
e_n, \ldots \quad $ be an orthonormed basis in $E$ and
$f_1,\ldots, f_n, \ldots \quad $ be an orthonormed basis in
$F$.  If

$$ae_m =\sum_p \alpha_{pm}e_p , \quad {\rm and} \quad
af_n =\sum_q \beta_{qn}f_q +
      \sum_r \gamma_{rn}e_r , $$

\noindent
then, since $(ae_m,f_n)=(e_m,af_n)$, one immediately
derives that $\gamma_{rn}=0$ for all values of $r$ and $n$.
Hence $F$ is also an invariant subspace. Thus, the
orthogonal compliment to each invariant subspace is also an
invariant subspace and therefore $W$ is completely
reducible.  In view of this result the problem reduces to
the determination of all irreducible $osp(3/2)$ modules
satisfying (i).

The position and the momentum operators are Hermitian
operators (hence also the Hamiltonian $H$, the square of
the angular momentum  ${\bf M}^2$ and its projections
$M_1$, $M_2$, $M_3$ are Hermitian operators) if and only

$$(a_k^-)^\dagger = a_k^+, \;\; k=1,2,3 \quad {\rm or \;
equivalently \; if}
\quad (c_p^-)^\dagger = c_p^+, \;\; p=0,1, \eqno(26) $$

\noindent
where $(a)^\dagger $ is the Hermitian conjugate to the
operator $a$.

As in the canonical case, one shows that the energy of any
such oscillator should be nonnegative.  Indeed, if $\psi$ is a
normed eigenvector of the Hamiltonian, $H\psi=E\psi$,
$(\psi,\psi)=1$, then, since $p_i$ and $r_i$ are Hermitian,
from (3) one has

$$E=(\psi,H\psi)=\sum_{i=1}^3[{1\over 2m}(p_i\psi,p_i\psi)+
{m\omega^2\over 2}(r_i\psi,r_i\psi)]>0.   $$

\smallskip
If $\psi_0$ is a state corresponding to the ground energy $E_0$
then

$$a_k^-\psi_0=0, \quad k=1,2,3, \eqno(27) $$

\noindent
since otherwise $a_k^-\psi_0$ would correspond to a state
with an energy $E_0-\omega \hbar$.

The irreducible representations (irreps), for which eqs.
(26) and (27) hold (we refer to them as to oscillator
representations) have been classified (among several
others) by Van der Jeugt [10]. They are star irreps [11]
with a highest weight.  Recently all such irreps have been
explicitly constructed [12]. They are infinite-dimensional.
Thus, the condition (i) and hence also all conditions
(i)-(iii) are satisfied with the oscillator representations
of the LS $osp(3/2)$, which we now proceed to describe.

The oscillator representations are labelled with the set of
all possible pairs $(p, q)$, where $p$ is an arbitrary
nonnegative half-integer, $p=0, 1/2, 1, 3/2, \ldots \quad$
and $q$ is any negative real number, such that $p+2q \leq
0$. The $osp(3/2)$ module corresponding to such a pair is
denoted as $W(p,q)$. Each such module is a direct sum of no
more than 8 irreducible infinite-dimensional modules $V(p,
q;M, J)$ of the even subalgebra $so(3)\oplus sp(2)$. $M$ is
the angular momentum of the oscillator in a state $\psi$
from $V(p, q;M, J)$,

$${\bf M}^2\psi \equiv [(M_1)^2+(M_2)^2+(M_3)^2]\psi =
M(M+1)\psi \quad \forall \psi \in V(p, q;M, J). \eqno(28)$$

\noindent
and $J$ is the analog of $M$ for the subalgebra $sp(2)$,
sitting in the right lower corner of matrix representations
(10).  More precisely, let

$$J^\pm \equiv J_1 \pm iJ_2=\mp {1 \over 2}(c_1^\mp)^2,
\quad J_3=-{1 \over 4}\{c_1^+,c_1^-\}. \eqno(29) $$

\noindent
Then $J^+$ and $J^-$ are the positive and
negative root vectors of $sp(2)=sl(2)$; $J_3$ is the Cartan
generator,

$$[J_3,J^\pm]=\pm J^\pm,\quad [J^+,J^-]=2J_3. \eqno(30)  $$

\noindent
Eqs. (29) give the usual realization of $sl(2)$ in terms of
para-Bose (and hence also in terms of Bose) operators. The
label $J$ is the "spin" of the reducible $sp(2)$ module
$V(p, q;M, J)$,

$${\bf J}^2\psi \equiv [(J_1)^2+(J_2)^2+(J_3)^2]\psi =
J(J+1)\psi \quad \forall \psi \in V(p, q;M, J). \eqno(31)$$

Let
$$\theta (x)=\cases {0, & for $ x<0$;\cr
                     1, & for $x \geq 0$.\cr } $$
Then the decomposition of the irreducible $osp(3/2)$
module $W(p,q)$ into a direct sum of irreducible
$so(3)\oplus sp(2)$ modules $V(p, q;M, J)$ reads [12]

$$\vcenter{\openup3\jot \halign{$#$ \hfil  \cr
W(p,q)=V(p,q;p,q) \oplus \theta (p-1)V(p,q;p-1,q-{1 \over 2})
\oplus \theta (p-1)V(p,q;p-1,q-1) \cr
\oplus \theta (p-{1\over 2})V(p,q;p,q-{1 \over 2})
\oplus (p+2q)[ \theta (p-{1\over 2})V(p,q;p,q-1)
\oplus V(p,q;p,q-{3 \over 2}) \cr
\oplus V(p,q;p+1,q-{1 \over 2})
\oplus V(p,q;p+1,q-1)].\cr
}} \eqno(32)$$

\noindent
The multipliers $p+2q$, $\theta (p-{1\over 2})$ and
$\theta (p-1)$  are to indicate that at certain
values of $p$ and $q$ some of the terms in the r.h.s. of
(32) are not present. For instance at $p+2q=0$ the last 4
terms in (32) disapear. There can be even less terms if in
addition $p=0$ or $p={1 \over 2}$. Observe that the $sp(2)$
"spin" $J$ of each $V(p, q;M, J)$ in (32) takes only
negative values. It corresponds  to unitarizable
infinite-dimensional representations of the noncompact real
form $su(1,1)$ of $sp(2)$.

The basis (${Z \!\!\! Z}_+$=all nonnegative integers)

$$|p,q;M,J;m,j>, \quad m=-M,-M+1,\ldots, M-1,M,
\quad j=J-n, \;\; n\in {Z \!\!\! Z}_+ \eqno(33) $$

\noindent
in $V(p, q;M, J)$ consists of eigenvectors of the Cartan
subalgebra, which is a linear span of $M_3$ and $J_3$. The
transformation of the basis under the action of all
$osp(3/2)$ generators is completely defined from its
transformation under the action of the Lie-super triple system
generators $c_0^\pm$ and $c_1^\pm$. The expressions for the
even generators, namely the para-Fermi generators
$c_0^\pm$, are simple:

$$c_0^\pm |p,q;M,J;m,j>=-|(M\mp m)(M\pm m+1)|^{1/2}
|p,q;M,J;m\pm 1,j>. \eqno(34) $$

The transformations under the action of the odd generators
$c_1^\pm$, i.e., the para-Bose operators, are more
involved. They follow and are in fact simpler then the
expressions derived in [12]:

$$\eqalignno{
{\rm For} \;p \geq 0  & \; {\rm and}\; p+2q\leq 0 & \cr
&c_1^\mp |p,q;p,q;m,j>
=\theta(p-{1\over 2}) m
\left|{(2q-1)(\pm q-j)\over
qp(p+1)}\right|^{1/2}
|p,q;p,q-{1\over 2};m,j\pm {1\over 2} > & \cr
& + \left|{(p+2q)(\pm q-j)(p+m+1)(p-m+1)\over
q(p+1)(2p+1)}\right|^{1/2}
|p,q;p+1,q-{1\over 2};m,j\pm {1\over 2} > &  \cr
& +\theta(p-1)\left|{(p-2q+1)(\pm q-j)(p-m)(p+m)\over
qp(2p+1)}\right|^{1/2}
|p,q;p-1,q-{1\over 2};m,j\pm {1\over 2}>&(35) \cr 
&&\cr
{\rm For} \;p \geq 0  & \; {\rm and}\; p+2q < 0 & \cr
& c_1^\mp |p,q;p+1,q-{1\over 2};m,j>=
-{m \over p+1}\left|\pm q-j \mp {1\over 2}\right|^{1/2}
|p,q;p+1,q-1;m,j\pm {1\over 2} > & \cr
&-{1\over p+1}\left|{p(p-2q+1)(\pm q-j\mp {1\over 2})
(p+m+1)(p-m+1)\over q(2p+1)}\right|^{1/2}
|p,q;p,q-1;m,j\pm {1\over 2} > &  \cr
&+\left|{(p+2q)(\mp q-j\mp{1\over 2})(p+m+1)(p-m+1)\over
q(p+1)(2p+1)}\right|^{1/2}
|p,q;p,q;m,j\pm {1\over 2} > & (36) \cr 
&&\cr
{\rm For} \;p > 0 & \; {\rm and}\; p+2q\leq 0 & \cr
& c_1^\mp |p,q;p,q-{1\over 2};m,j>=m
\left|{(2q-1)(\mp q-j\mp{1\over 2})\over
qp(p+1)}\right|^{1/2}
|p,q;p,q;m,j \pm {1\over 2} > & \cr
&+{1 \over p+1}
\left|{2p(p+2q)(\pm q-j \mp {1\over 2})(p+m+1)(p-m+1)\over
(2q-1)(2p+1)}\right|^{1/2}
|p,q;p+1,q-1;m,j\pm {1\over 2} > &  \cr
&-{1\over p}\left|{2(p+1)(p-2q+1)(\pm q-j\mp {1\over 2})(p+m)(p-m)\over
(2q-1)(2p+1)}\right|^{1/2}
|p,q;p-1,q-1;m,j\pm {1\over 2} >  & \cr
&+{m \over p(p+1)}\left|{(p-2q+1)(p+2q)(\pm q-j\mp {1\over
2}) \over q(2q-1)}\right|^{1/2}
|p,q;p,q-1;m,j\pm {1\over 2}>,  &(37) \cr 
&&\cr
{\rm For} \;p \geq 1 & \; {\rm and}\; p+2q\leq 0 & \cr
& c_1^\mp |p,q;p-1,q-{1\over 2};m,j>=
{m \over p}\left|2(\pm q-j \mp {1 \over 2}\right|^{1/2}
|p,q;p-1,q-1;m,j\pm {1\over 2} > & \cr
&+\left|{(p-2q+1)(\mp q-j \mp {1 \over 2})(p-m)(p+m)\over
qp(2p+1)}\right|^{1/2}
|p,q;p,q;m,j\pm {1\over 2} > &  \cr
&+{1 \over p}\left|{(p+2q)(p+1)(\pm q-j \mp {1 \over 2})
(p+m)(p-m)\over
q(2p+1)}\right|^{1/2}
|p,q;p,q-1;m,j\pm {1\over 2} >,& (38)\cr 
&&\cr
{\rm For} \;p \geq 0 & \; {\rm and}\; p+2q < 0 & \cr
&c_1^\mp |p,q;p+1,q-1;m,j>=
-{m \over p+1}
\left|2(\mp q-j)\right|^{1\over 2}
|p,q;p+1,q-{1\over 2};m,j\pm {1\over 2} > & \cr
&+{1 \over p+1}\left|{2p(p+2q)(\mp q-j)(p+m+1)(p-m+1)\over
(2q-1)(2p+1)}\right|^{1/2}
|p,q;p,q-{1\over 2};m,j\pm {1\over 2} > & \cr
&+\left|{2(p-2q+1)(\pm q-j\mp 1)(p+m+1)(p-m+1)
\over (2q-1)(p+1)(2p+1)}\right|^{1/2}
|p,q;p,q-{3 \over 2};m,j\pm {1\over 2}>& (39) \cr 
&&\cr
{\rm For} \;p > 0 & \; {\rm and}\; p+2q < 0 & \cr
& c_1^\mp |p,q;p,q-1;m,j>=
-2m\left|{q(\pm q-j \mp 1)\over
p(p+1)(2q-1)}\right|^{1/2}
|p,q;p,q-{3\over 2};m,j\pm {1\over 2} > & \cr
&+{m\over p(p+1)}\left|{(p-2q+1)(p+2q)(\mp q-j)
\over q(2q-1)}\right|^{1/2}
|p,q;p,q-{1\over 2};m,j\pm {1\over 2}> & \cr
&-{1\over p+1}\left|{p(p-2q+1)(\mp q-j)(p+m+1)(p-m+1)
\over q(2p+1)}\right|^{1/2}
|p,q;p+1,q-{1\over 2};m,j\pm {1\over 2} > &  \cr
&+{1\over p}\left|{(p+2q)(p+1)(\mp q-j)(p+m)(p-m)\over
q(2p+1)}\right|^{1/2}
|p,q;p-1,q-{1\over 2};m,j\pm {1\over 2}>,& (40) \cr 
&&\cr
{\rm For} \;p \geq 1 & \; {\rm and}\; p+2q\leq 0 & \cr
& c_1^\mp |p,q;p-1,q-1;m,j>=
{m \over p}
\left|2(\mp q-j)\right|^{1\over 2}
|p,q;p-1,q-{1\over 2};m,j\pm {1\over 2} > & \cr
&-{1 \over p}\left|{2(p-2q+1)(p+1)(\mp q-j)(p+m)(p-m)\over
(2q-1)(2p+1)}\right|^{1/2}
|p,q;p,q-{1\over 2};m,j\pm {1\over 2} > & \cr
&+\left|{2(p+2q)(\pm q-j\mp 1)(p+m)(p-m)
\over p(2q-1)(2p+1)}\right|^{1/2}
|p,q;p,q-{3 \over 2};m,j\pm {1\over 2}>,& (41) \cr 
&&\cr
{\rm For} \;p \geq 0 & \; {\rm and}\; p+2q < 0 & \cr
& c_1^\mp |p,q;p,q-{3\over 2};m,j>=
-\theta (p-{1\over 2})2m \left|{q(\mp q-j\pm {1\over 2})\over
p(p+1)(2q-1)}\right|^{1/2}
|p,q;p,q-1;m,j \pm {1\over 2} > & \cr
&+\left|{2(p-2q+1)(\mp q-j \pm {1\over 2})(p+m+1)(p-m+1)\over
(2q-1)(p+1)(2p+1)}\right|^{1/2}
|p,q;p+1,q-1;m,j\pm {1\over 2} > &  \cr
&\theta (p-1)\left|{2(p+2q)(\mp q-j\pm {1\over 2})(p+m)(p-m)\over
p(2q-1)(2p+1)}\right|^{1/2}
|p,q;p-1,q-1;m,j\pm {1\over 2} >  & (42) \cr 
}$$

The above eqs. (34)-(42) are not easy to be derived.  It is
even quite difficult to verify that they give a
representation of the Lie-super triple relations (16). In
doing so and, more generally, applying eqs. (35)-(42) one
should have in mind that whenever a $\theta$-function in
front of a certain  term in the r.h.s. vanishes, then the
corresponding term should be canceled out independently of
the fact that some other multipliers in the same term could
be undefined (at $p=0$ one has sometimes factors $0 \over
0$).

The transformation relations of the basis under the action
of some even generators, which follow from (33)-(42), read:

$$\eqalignno{
& so(3)\; {\rm generators}\; M^\pm=M_1\pm i M_2,\;M_3: &  \cr
&M^\pm |p,q;M,J;m,j>=|(M\mp m)(M\pm m+1)|^{1/2}
|p,q;M,J;m\pm 1,j>,  & (43)\cr
&M_3|p,q;M,J;m,j>=m|p,q;M,J;m,j>, & (44)  \cr
&{\bf M}^2|p,q;M,J;m,j>=M(M+1)|p,q;M,J;m,j>; & (45) \cr
&&\cr
&sp(2)\; {\rm generators}: & \cr
&J^\pm |p,q;M,J;m,j>=|(J\pm j+1)(J\mp j)|^{1/2}
|p,q;M,J;m,j\pm 1>, & (46) \cr
&J_3|p,q;M,J;m,j>=j|p,q;M,J;m,j>. & (47) \cr
&{\bf J}^2|p,q;M,J;m,j>=J(J+1)|p,q;M,J;m,j>; & (48) \cr
}$$

Postulate that the basis (33) within each $so(3)\oplus sp(2)$
module $V(p, q;M, J)$ is orthonormed and that the different
such modules are orthogonal to each other. With respect to
this metric
the hermiticity conditions (26) hold.
Consequently also  $(r_1, r_2, r_3)$,
$(p_1, p_2, p_3)$, ${\bf M}^2$, $(M_1, M_2, M_3)$ and $H$ are
Hermitian operators. Thus the condition (i) and hence all
quantum conditions (i)-(iii) are satisfied within any
oscillator module $W(p,q)$, considered as a state space of
a noncanonical oscillator. Hence any triple
({\bf r, p}, $W(p,q)$) is a Wigner quantum oscillator.

\vskip 8mm
\noindent
{\bf 4. Energy and  angular momentum spectrum of the oscillators.
Multiplicities}

\bigskip
\noindent
The Hamiltonian (3) of the $osp(3/2)$ oscillator is
proportional to $J_3$ (see (29)),

$$H={\omega \hbar \over 2}\{c_1^+,c_1^-\}=
-2\omega \hbar J_3 \in sp(2) . \eqno(49) $$

The operators $H$, ${\bf M}^2$ and $M_3$ commute with each
other. According to (44), (45), (47) and (49) the basis
(33) is diagonal with respect to these operators. In
particular

$$H|p,q;M,J;m,j>=-2\omega \hbar j|p,q;M,J;m,j>. \eqno(50) $$

Each state space $W(p,q)$ contains a subspace $V(p,q;M,J)$ with
$J=q$ and at least one subspace with $J=q-{1\over 2}$.
Taking into account that within $V(p,q;M,J)$
$j=J,J-1,J-2,\ldots \;$ from (50) one obtains the spectrum
$E_n$ of $H$ in the subspaces with $J=q$ and
$J=q-{1\over 2}$ :

$${\rm In}\; V(p,q;p,q) \quad E_n=\omega\hbar (2n-2q),\quad
n=0,1,2,3,\ldots \; ;\eqno(51)$$
$${\rm In}\; V(p,q;M,q-{1\over 2}) \quad
E_n=\omega\hbar [(2n+1)-2q],\quad n=0,1,2,3,\ldots \;
.\eqno(52)$$

\noindent
Combining (51) and (52) one obtains

$$E_n=\omega\hbar (n-2q),\quad n=0,1,2,3,\ldots \;. \eqno(53) $$

The energies of the states in the other subspaces
$V(p,q;M,J)$ do not change the spectrum (53); they change
only the multiplicities of the energy levels. Therefore eq.
(53) gives the energy levels of the oscillator in a state
space $W(p,q)$

{}From the decomposition (32) and eq. (28) one concludes that
the angular momentum of an $osp(3/2)$ oscillator in a state
space $W(p,q)$ can take at most three different values,
namely $$M=p-1,p,p+1.$$ The states from a subspace
$V(p,q;M,J)$ carry an angular momentum $M$. Each such space
is infinite-dimensional. Hence the multiplicity of each
allowed value $M$ of the angular momentum is also
infinite-dimensional.

In order to analyze the multiplicities of the stationary
states, we first observe that each irreducible $so(3)\oplus
sp(2)$ module $V(p,q;M,J)$ is a tensor product of an
irreducible $2M+1$ dimensional $so(3)$ module $[M]$ with an
irreducible infinite-dimensional $sp(2)$ module $[J]$,

$$V(p,q;M,J)=[M]\otimes [J],\eqno(54) $$

\noindent
An operator $a$ from $so(3)$  acts in $[M]\otimes [J]$ as
$a\otimes id$, whereas an operator $b$ from $sp(2)$ acts as
$id \otimes b$ ($id$=identity operator).
Each basis vector $|p,q;M,J;m,j>\in V(p,q;M,J) $ can be
represented as

$$|p,q;M,J;m,j>=|M,m>\otimes |J,j>. \eqno(55)$$

\noindent
Then

$$\eqalignno{
&{\bf M}^2|M,m>=M(M+1)|M,m>,\quad
M_3|M,m>=m|M,m>,&(56) \cr
&{\bf J}^2|J,j>=J(J+1)|J,j>,\quad
J_3|J,j>=j|J,j>.& (57) \cr
}$$

{}From (54) it is evident that the linear envelope of all
$|p,q;M,J;m,j>$ states from $V(p,q;M,J)$ with a fixed $j$,
namely with a fixed energy, is an irreducible $(2M+1)$
dimensional $so(3)$ module $[M]$. This observation together
with the decomposition (32) leads to the conclusion that
the eigenspace $V(E_n)$ of the Hamiltonian $H$ in $W(p,q)$
is  generally reducible $so(3)$ module.

Let

$$\varphi (x)=\cases {1, & for $ x=0,1,2,\ldots $;\cr
                   0, & otherwise.\cr } \eqno(58)  $$

Then for the eigenspace $V(E_n)$ of the Hamiltonian $H$ in
$W(p,q)$ we obtain:

$$\eqalignno{
&V(E_n)=\theta(p-1)\theta(n-1)[p-1]
\oplus \{1-\theta(p+2q)\}\theta(n-1)[p+1]
\oplus \{\theta(p-{1\over 2})\varphi({n-1\over 2})
&\cr
&+\varphi({n\over 2})+\{1-\theta(p+2q)\}\theta(p-{1\over 2})
\varphi({n\over 2}-1)+\{1-\theta(p+2q)\}
\varphi({n-3\over 2})\}[p].& (59)\cr
}$$

{}From eq. (59) one concludes that an oscillator with a state
space $W(p,q)$ has

\smallskip
\noindent
a) $\theta(p-1)\theta(n-1)(2p-1)$ states with an angular
momentum $M=p-1$ and energy $E_n$,

\smallskip
\noindent
b) $\{\theta(p-{1\over 2})\varphi({n-1\over 2})
+\varphi({n\over 2})+\{1-\theta(p+2q)\}\theta(p-{1\over 2})
\varphi({n\over 2}-1)+\{1-\theta(p+2q)\}
\varphi({n-3\over 2})\}(2p+1)$ states with an angular
momentum $M=p$ and energy $E_n$,

\smallskip
\noindent
c) $ \{1-\theta(p+2q)\}\theta(n-1)(2p+3)$ states with an angular
momentum $M=p+1$ and energy $E_n$,

\smallskip
\noindent
Summing up the expressions in a), b) and c) one obtains
the number of the (linearly independent) states with energy
$E_n$, i.e., $dimV(E_n)$.

Let us consider the cases corresponding to different values
of $p$.

\bigskip
{\bf Oscillators with p=0}
\smallskip
Since $q<0$, then also $p+2q<0$ and according to (59)

$$V(E_0)=[0],\quad V(E_1)=[1],\quad  V(E_n)=[0]\oplus [1],\;
n>1. \eqno(60) $$

In particular the ground state ($n=0$) is nondegenerated and
carries an angular momentum zero. Depending on the value of
$q$ the energy of the ground state $E_0=-2\omega\hbar q$
can be arbitrarily close to zero, but never zero. The first
excited states ($n=1$) have $M=1$. In all other cases
($n>1$) the eigenspace of the Hamiltonian is reducible with
respect to $so(3)$. There is one state with angular
momentum zero and three spaces with $M=1$.

\bigskip
{\bf An oscillator with p=1/2 and p+2q=0}
\smallskip

This case corresponds to the oscillator considered in [1].
The angular momentum of each state is $M={1\over 2}$,
$V(E_n)=[{1\over 2}]$ for any  $n$; there are two states
corresponding to every energy level. The state space
$W({1\over 2}, -{1\over 4})$ is an infinite direct sum of
2-dimensional representations of $so(3)$. In this case the
expressions (35)-(42) are greatly simplified; $c_0^\pm$ are
usual Fermi operators, whereas $c_1^\pm$ are Bose operators.
Moreover the bosons anticommute with the fermions. The
representation of $osp(3/2)$  is one of the
metaplectic representations of the superalgebra [10].
The energy spectrum of the oscillator is the same as for an
one-dimensional canonical harmonic oscillator,

$$E_n=\omega\hbar (n+{1\over 2}),\quad n=0,1,2,\ldots \eqno(61)  $$

\bigskip
{\bf Oscillators with p=1/2 and p+2q$<$0}
\smallskip
The subspace corresponding to the ground energy is
two-dimensional and carries an angular momentum $1\over 2$.
The angular momentum of all other states is either  $1\over
2$ or $3\over 2$. More precisely one has:

$$V(E_0)=\left[{1\over 2}\right],\quad
V(E_1)=\left[{1\over 2}\right]\oplus \left[{3\over 2}\right],
\quad  V(E_n)=\left[{1\over 2}\right]\oplus \left[{1\over 2}\right]\oplus
\left[{3\over 2}\right], \; n>1. \eqno(62) $$

\bigskip
{\bf Oscillators with p$>$1/2 and p+2q=0}
\smallskip
The angular momentum of the ground subspace is $M=p$. There
are $2p+1$ ground states; all other states have an angular
momentum $p$ or $p-1$:

$$V(E_0)=[p],\quad V(E_n)=[p]\oplus [p-1], \; n\geq 1. \eqno(63) $$

\bigskip
{\bf Oscillators with p$>$1/2 and p+2q$<$0}
\smallskip
The structure of the ground subspace is the same as in the
previous case. The angular momentum of any other eigenspace
of the Hamiltonian is a reducible $so(3)$ module with
angular momentum $M=p-1,p,p+1$. Its $so(3)$ content reads:

$$V(E_0)=[p],\quad V(E_1)=[p-1] \oplus [p] \oplus [p+1],
\quad  V(E_n)=[p-1] \oplus [p]\oplus [p] \oplus [p+1], \;
n>1. \eqno(64) $$

\vskip 8mm
\noindent
{\bf 5. Discussions}

\bigskip
\noindent
{}From the above considerations it is clear that
the $osp(3/2)$ oscillators differ essentially from
the canonical 3-dimensional oscillator. We underline
some of the main points in this respect.

The coordinates of any $osp(3/2)$ oscillator
anticommute with each other (see eq.(22)). Therefore one
cannot have a coordinate (or $x$-) representation for the
wave function.  The geometry of the oscillator is
noncommutative. For the same reason there exists no
momentum representation. Here we have considered the case
with $H$, ${\bf M}^2$ and $M_3$ being simultaneously
diagonal, namely an energy - angular momentum
representation.

The canonical oscillator can be in a state with any integer
value of the angular momentum $M$, but never in a state with
half-integer values of $M$.  An $osp(3/2)$ oscillator
allows at most three values of the angular momentum,
$M=p-1,\;p,\;p+1$, but they can be either  integers or
half-integers. In particular if $p={1\over 2}$ then the
angular momentum takes only one value $M={1\over 2}$; if
$p=0$ then $M=0,1$; in all other cases $M$ can have three
different values as indicated above.

The energy spectrum of any $osp(3/2)$ oscillator is
equidistant. In four cases, namely in the state spaces
$W(p,-{3\over 4})$ with $p=0,{1\over 2},1,{3\over 2}$ the
spectrum is the same as of the canonical 3-dimensional
oscillator,

$$E_n=\omega\hbar (n+{3\over 2}),\quad n=1,2,3,\ldots \;.
\eqno(65) $$

In all other cases the spectrum is different. It may be even
very different for large values of $p$. Indeed the
conditions $p+2q\leq 0$, $q<0$ put restrictions from bellow
for the ground energy, namely $E_0\geq \omega\hbar p$.
Even for small values of $p$, but large values  of $q$
the ground energy $E_0 $ may be much above the canonical ground
energy ${3\over 2}\omega\hbar $.

Another essential new feature we like to point out is the
degeneracy of the ground states. As one can see from eq.
(60), the eigenspace $V(E_0)$ of the Hamiltonian is
nondegenerated only in the state spaces $W(0,q)$, i.e.,
those with $p=0$. In all cases the states from $V(E_0)$
carry one and the same angular momentum, namely $M=p$ if
$V(E_0)\subset W(p,q)$; the ground subspace transforms as an
irreducible $so(3)$ module $V(E_0)=[p]$.  In the state
spaces with $p\ge 1$ the eigenspaces $V(E_n),\;n>0,$ of $H$
carry different angular momentum, they are reducible with
respect to $so(3)$.

As a last remark we mention that all our considerations
are in the Heisenberg picture; the operators are generally
time dependent. The only time independent operators (from
those we have considered) are the Hamiltonian $H$, the
square of the angular momentum operator ${\bf M}^2$ and its
projections $(M_1, M_2, M_3)$, namely the operators
generating the stability subalgebra $so(3)\oplus gl(1)$.
The root vectors $J^\pm$ of $sp(2)$ and all odd generators
(see (8) or (21)) depend on time.

\vskip 24pt
\noindent
{\bf Acknowledgements}
\vskip 12pt
One of us (T. P.) is thankful to Prof. H. D. Doebner for
the kind hospitality at the Arnold Sommerfeld Institute for
Mathematical Physics, where some of the results in the
present investigation have been obtained.

\vskip 24pt
\noindent
{\bf References}

\vskip 12pt
\settabs \+  [11] & I. Patera, T. D. Palev, Theoretical
   interpretation of the experiments on the elastic \cr

\+ [1] & Palev T D and Stoilova N I 1994 {\it J. Phys. A: Math.
         Gen.} {\bf 27} 977  \cr

\+ [2] & Wigner E P 1950 {\it Phys. Rev.} {\bf 77} 711 \cr

\+ [3] & Ohnuki Y and Kamefuchi S  1978 {\it J. Math. Phys.}
         {\bf 19} 67; 1979 {\it Z. Phys. C} {\bf 2} 367 \cr

\+     & Okubo S 1980 {\it Phys. Rev. D} {\bf 22} 919;\cr

\+     & Mukunda N, Sudarshan E C G, Sharma J K and
         Mehta C L 1980 {\it J. Math. Phys.} {\bf 21} 2386;\cr

\+     & Ohnuki Y and Watanabe S 1992 {\it J. Math. Phys.}
         {\bf 33} 3653 \cr

\+ [4] & Palev T D 1982  {\it J. Math. Phys.} {\bf 23} 1778;
         1982 {\it Czech. J. Phys. B} {\bf 3} 680 \cr

\+     & Kamupingene A H, Palev T D and Tsaneva S P 1986
         {\it J. Math. Phys.} {\bf 27} 2067 \cr

\+ [5] & Kac V G 1977 {\it Adv. Math.} {\bf 26} 8 \cr

\+ [6] & Palev T D 1982 {\it J. Math. Phys.} {\bf 23} 1100 \cr

\+ [7] & Okubo S 1993 {\it University of Rochester preprint}
         ER-1335, ERO40685-784 \cr

\+ [8] & Green H S 1953 {\it Phys. Rev.} {\bf 90} 270 \cr

\+ [9] & Macfarlane A J 1994 {\it J. Math. Phys.} {\bf 35}
         1054 \cr

\+ [10] & Van der Jeugt J 1984 {\it J. Math. Phys.} {\bf
          25} 3334  \cr

\+ [11] & Scheunert M, Nahm W and Rittenberg V 1977
          {\it J. Math. Phys.} {\bf 18} 146 \cr

\+ [12] & Ky N A, Palev T D and Stoilova N I 1992
          {\it J. Math. Phys.} {\bf 33} 1841 \cr

\end